\documentclass[3p,times,twocolumn]{elsarticle}

\usepackage{ecrc}
\usepackage{booktabs}
\usepackage{epstopdf}
\usepackage{xcolor}
\usepackage[percent]{overpic}

\usepackage{SIunits}
\usepackage{upgreek,xspace}
\def\textalpha{\ensuremath\upalpha}
\def\textbeta{\ensuremath\upbeta}
\def\textgamma{\ensuremath\upgamma}
\def\textsigma{\ensuremath\upsigma}
\def\textLambda{\ensuremath\Uplambda}
\def\textpi{\ensuremath\uppi}

\volume{00}

\firstpage{1}

\journalname{Nuclear Physics B Proceedings Supplement}

\runauth{M. Ku\'zniak (DEAP Collaboration)}

\jid{nuphbp}

\jnltitlelogo{Nuclear Physics B Proceedings Supplement}

\usepackage{lineno}

\usepackage[figuresright]{rotating}

\newcommand{\fprompt}{F$_{\mathrm{prompt}}$}

\newcommand{\kevee}{keV$_{\mathrm{ee}}$}

\begin{document}

\begin{frontmatter}



\dochead{}

\title{DEAP-3600 Dark Matter Search}

\author[TRIUMF]{P.-A.~Amaudruz}
\author[LU]{M.~Batygov}
\author[UofA]{B.~Beltran}
\author[QU]{J.~Bonatt}
\author[QU]{M.\,G.~Boulay}
\author[QU]{B.~Broerman}
\author[UofA]{J.\,F.~Bueno}
\author[RHUL]{A.~Butcher}
\author[QU]{B.~Cai}
\author[QU]{M.~Chen}
\author[UofA]{R.~Chouinard}
\author[LU]{B.\,T.~Cleveland}
\author[QU]{K.~Dering}
\author[SNOLAB]{J.~DiGioseffo}
\author[LU]{F.~Duncan}
\author[SNOLAB]{T.~Flower}
\author[LU]{R.~Ford}
\author[QU]{P.~Giampa}
\author[UofA]{P.~Gorel}
\author[Carleton]{K.~Graham}
\author[UofA]{D.\,R.~Grant}
\author[TRIUMF]{E.~Guliyev}
\author[UofA]{A.\,L.~Hallin}
\author[QU]{M.~Hamstra}
\author[QU]{P.~Harvey}
\author[LU]{C.\,J.~Jillings}
\author[QU]{M.~Ku\'zniak\corref{cor1}}
\ead{kuzniak@owl.phy.queensu.ca}
\cortext[cor1]{Corresponding author.}
\author[LU,SNOLAB]{I.~Lawson}
\author[LU,SNOLAB]{O.~Li}
\author[LU,SNOLAB]{P.~Liimatainen}
\author[RAL]{P.~Majewski}
\author[QU]{A.\,B.~McDonald}
\author[UofA]{T.~McElroy}
\author[LU,SNOLAB]{K.~McFarlane}
\author[RHUL]{J.~Monroe}
\author[TRIUMF]{A.~Muir}
\author[QU]{C.~Nantais}
\author[UofA]{C.~Ng}
\author[QU]{A.\,J.~Noble}
\author[Carleton]{C.~Ouellet}
\author[SNOLAB]{K.~Palladino}
\author[QU]{P.~Pasuthip}
\author[Sussex]{S.\,J.\,M.~Peeters}
\author[LU,SNOLAB]{T.~Pollmann}
\author[QU]{W.~Rau}
\author[TRIUMF]{F.~Reti\`ere}
\author[RHUL]{N.~Seeburn}
\author[UofA]{K.~Singhrao}
\author[QU]{P.~Skensved}
\author[TRIUMF]{B.~Smith}
\author[QU]{T.~Sonley}
\author[UofA]{J.~Tang}
\author[LU,SNOLAB]{E.~V\'azquez-J\'auregui}
\author[QU]{L.~Veloce}
\author[RHUL]{J.~Walding}
\author[QU]{M.~Ward}

\address{\em (DEAP Collaboration)\\}

\address[UofA]{Department of Physics, University of Alberta, Edmonton, Alberta, T6G 2R3, Canada}
\address[Carleton]{Department of Physics, Carleton University, Ottawa, Ontario, K1S 5B6, Canada}
\address[LU]{Department of Physics and Astronomy, Laurentian University, Sudbury, Ontario, P3E 2C6, Canada}
\address[QU]{Department of Physics, Engineering Physics, and Astronomy, Queen's University, Kingston, Ontario, K7L 3N6, Canada}
\address[RHUL]{Royal Holloway University London, Egham Hill, Egham, Surrey TW20 0EX, United Kingdom}
\address[RAL]{Rutherford Appleton Laboratory, Harwell Oxford, Didcot OX11 0QX, United Kingdom}
\address[SNOLAB]{SNOLAB, Lively, Ontario, P3Y 1M3, Canada}
\address[Sussex]{University of Sussex, Sussex House, Brighton, East Sussex BN1 9RH, United Kingdom}
\address[TRIUMF]{TRIUMF, Vancouver, British Columbia, V6T 2A3, Canada}

\begin{abstract}
The DEAP-3600 experiment is located 2 km underground at SNOLAB, in Sudbury, Ontario. It is a single-phase detector that searches for dark matter particle interactions within a 1000-kg fiducial mass target of liquid argon. A first generation prototype detector (DEAP-1) with a 7-kg liquid argon target mass demonstrated a high level of pulse-shape discrimination (PSD) for reducing \textbeta/\textgamma\ backgrounds and helped to develop low radioactivity techniques to mitigate surface-related \textalpha\ backgrounds. Construction of the DEAP-3600 detector is nearly complete and commissioning is starting in 2014. The target sensitivity to spin-independent scattering of Weakly Interacting Massive Particles (WIMPs) on nucleons of $10^{-46}$~cm$^2$ will allow one order of magnitude improvement in sensitivity over current searches at 100 GeV WIMP mass.
This paper presents an overview and status of the DEAP-3600 project and discusses plans for a future multi-tonne experiment, DEAP-50T.
\end{abstract}

\end{frontmatter}
\section{Introduction}\label{sec:intro}
In the standard model of cosmology (\textLambda CDM) a significant part of the matter present in the universe is so-called dark matter~\cite{WMAP, Planck}.
A currently favoured hypothesis is that this dark
matter is comprised of Weakly Interacting Massive Particles (WIMPs), which are not included in the standard model of particle physics (SM)
and have so far remained undetected.

The sensitivity of DEAP-3600~\cite{DEAPTaup2012,Gorel2014} and other experiments currently under construction will directly test predictions
of a number of theoretical SM extensions, such as supersymmetry (SUSY).
The two leading SUSY models, the cMSSM and NUHM, fit all of the currently available data including 
indirect searches, Planck results, recent SUSY searches and the Higgs discovery at the LHC as well as dark matter exclusions from direct detection 
experiments, with the leading upper limit from LUX~\cite{Akerib2014}.
To accommodate the global data, these models predict relatively heavy WIMPs,
with central values between a few hundred GeV and about 1~TeV~\cite{Roszkowski2014, Kowalska2013, Strege2013, Bechtle2012}.

The best-fit WIMP parameters in the cMSSM and NUHM models are within sensitivity
of the upcoming class of experiments (DEAP-3600 and XENON1T~\cite{Xenon1T}),
while possibly evading detection at the LHC, even for the future 14~TeV run because of the very high allowed superparticle masses.
A number of more general models, e.g. p9MSSM~\cite{Fowlie2013}, lead to similar conclusions, although the best-fit parameters can vary.

\section{The DEAP-3600 detector}
DEAP-3600, located at SNOLAB, will perform a dark matter particle search on liquid argon with sensitivity to the
spin-independent WIMP-nucleon cross-section of 10$^{-46}$~cm$^2$, a factor of approximately 20
increase over current experiments, as shown in Fig.~\ref{fig:deap-3600sensitivity}.
\begin{center}
\begin{figure}[htb]
\includegraphics[width=\columnwidth,trim=0 0 0.5cm 0,clip=true]{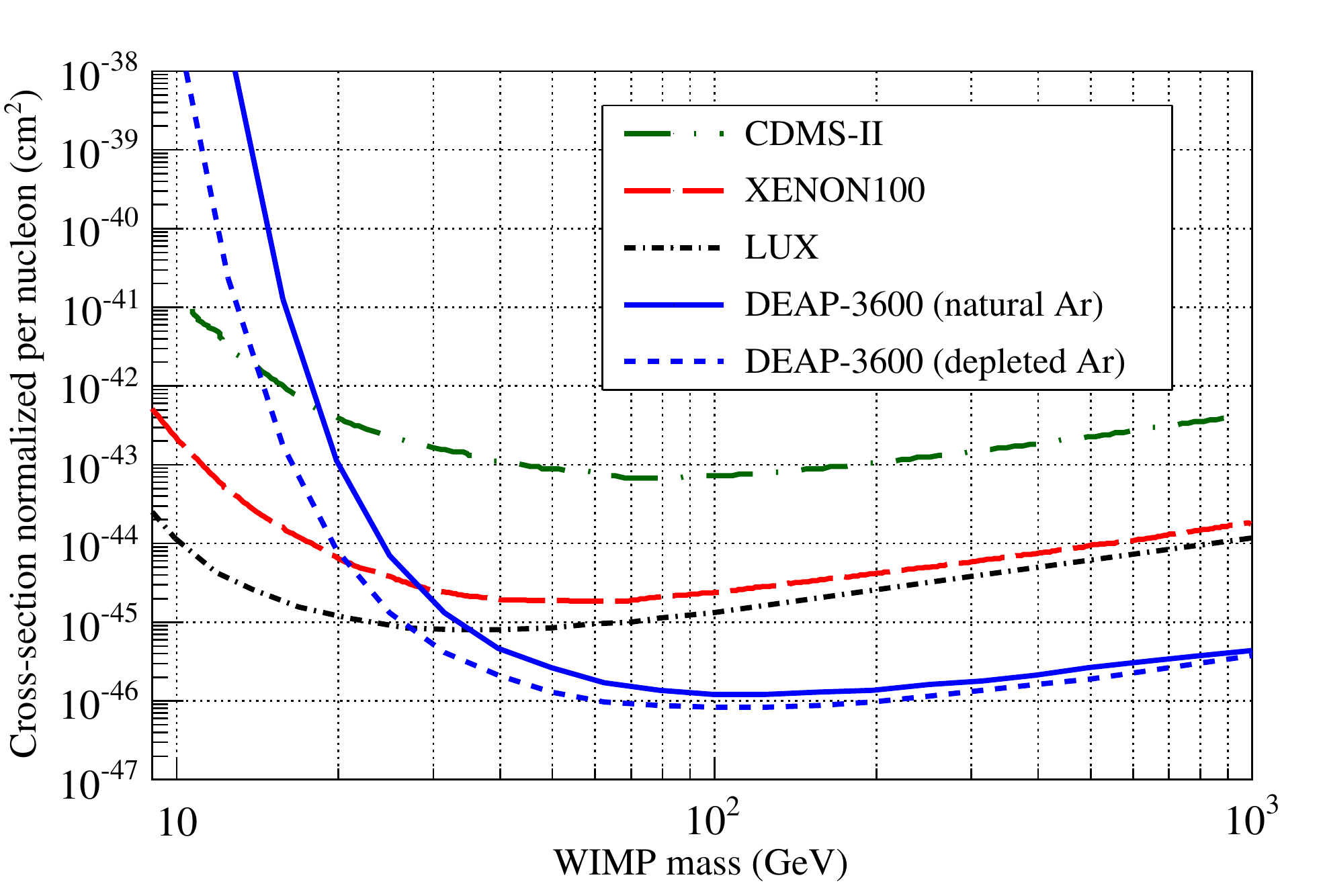}
\caption{Dark matter sensitivity of liquid argon expressed as a limit on spin-independent WIMP-nucleon scattering cross-section. Shown are the current experimental limits from the CDMS-II~\cite{cdms}, XENON100~\cite{Aprile2012}, and LUX~\cite{Akerib2014} detectors, and the expected sensitivity for 3 tonne-years of natural liquid argon with a 15~\kevee\ threshold, and with a 12~\kevee\ threshold for low-radioactivity argon (LRA) that has been depleted in $^{39}$Ar by a factor of 100.}
\label{fig:deap-3600sensitivity}
\end{figure}
\begin{figure}[h]
\includegraphics[width=\columnwidth,trim=0 0.9cm 0 1cm,clip=true]{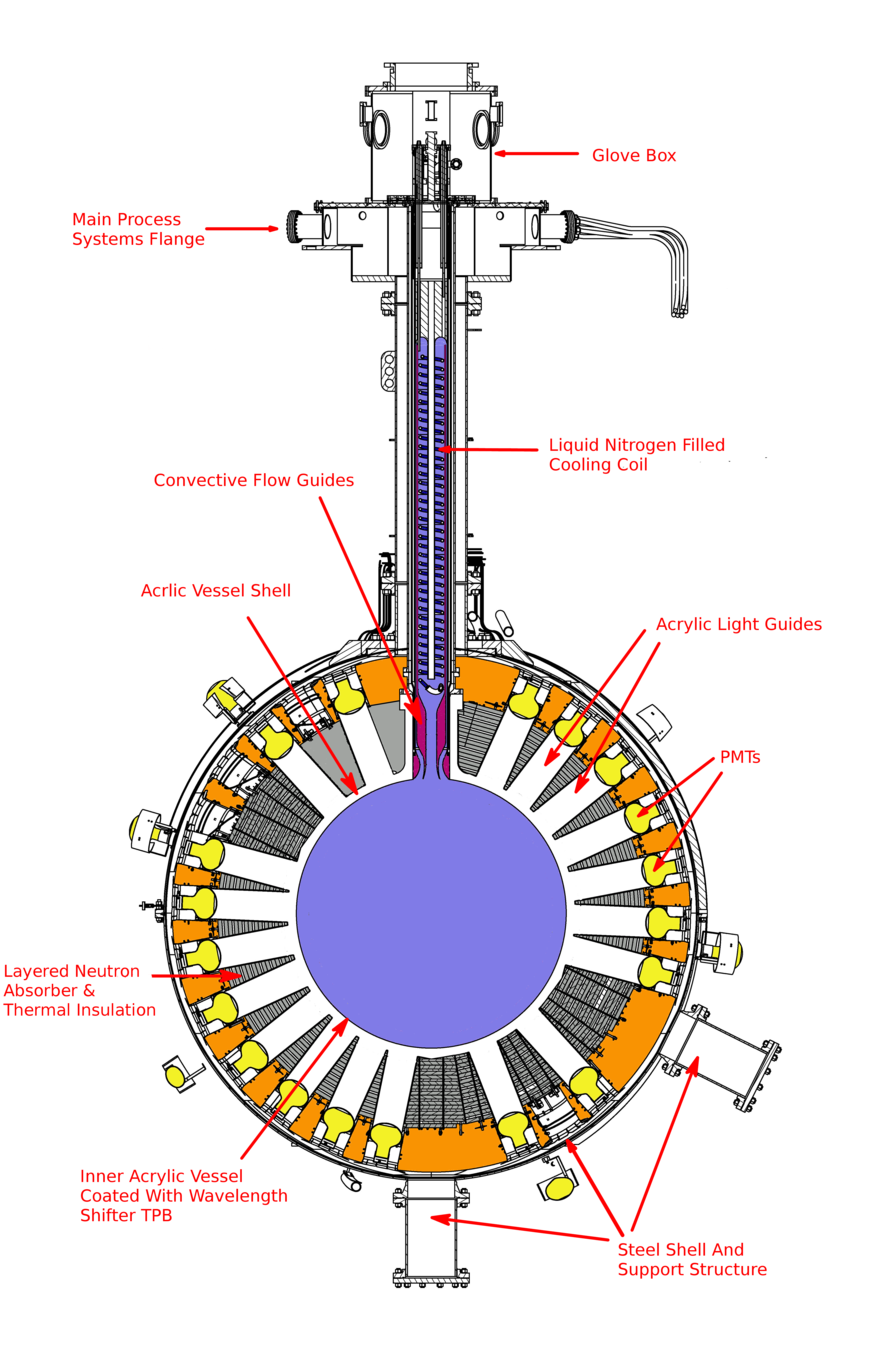}
\caption{The DEAP-3600 detector. The acrylic vessel has an inner radius of 85~cm and holds
3600~kg of liquid argon, which is viewed by 255 8-inch diameter high quantum efficiency Hamamatsu R5912-HQE PMTs
through 50-cm long lightguides.}
\label{fig:deap-3600schematic}
\end{figure}
\end{center}

The detector (Fig.~\ref{fig:deap-3600schematic}) is comprised of a large spherical volume of
natural liquid argon contained in a transparent acrylic vessel (AV) and
viewed by 255 photomultiplier tubes (PMT) that detect scintillation light generated in the 
argon target mass. The inner AV surface must be coated with a thin layer of wavelength shifter,
1,1,4,4-tetraphenyl-1,3-butadiene (TPB), to convert 128~nm argon scintillation light into visible blue light,
which is then efficiently transmitted through acrylic lightguides to the PMTs. A custom, large-area 
vacuum-deposition source~\cite{Pollmann2012}, developed by the DEAP collaboration, is used to deposit TPB 
uniformly over the approximately 10~m$^2$ acrylic surface before the detector is filled with cryogen.
The lightguides and polyethylene filler blocks provide neutron shielding
and thermal insulation between the cryogenic acrylic vessel and the much warmer PMTs.
The inner detector is housed in a large stainless-steel spherical shell, which itself is immersed in an
8~meter diameter ultrapure water tank and instrumented with PMTs, serving as a radiation shield and
\v{C}erenkov veto for cosmogenic muons.

The cryogenic requirements of DEAP are entirely met through the use of liquid nitrogen,
which is gravity fed from a large dewar on top of the experimental hall to the cooling coils in the
detector neck. Boil-off N$_2$ is collected and
returned to the dewar where it is re-condensed by cryocoolers. Liquid argon is held either
in a large storage dewar or inside the detector. The cryogenics operates in a stable closed loop
and is controlled by an industrial DeltaV system.

The data acquisition system was developed at TRIUMF and is described in detail elsewhere~\cite{Amaudruz2012,Retiere2012}.
In summary, PMT signals are decoupled from the high voltage by a set of custom analog signal-conditioning
boards, which split the PMT signals into three outputs: high-gain, low-gain, and a twelve
channel analog sum (ASUM). The ASUM is used by a custom FPGA-based Digitizer and Trigger Module to generate
trigger decisions depending on a pulse-shape discrimination (PSD) and selection.
High- and low-gain signals are digitized with commercial CAEN V1720 and V1740 digitizers, respectively, and handled
by the MIDAS DAQ system.

In order to reduce the risk of accidental contamination, we are planning to calibrate the detector using
a combination of external \textgamma/neutron sources, external and internal optical calibration systems,
and internal backgrounds, such as $^{39}$Ar events and residual surface backgrounds.

The water tank is equipped with three vertical calibration tubes that will allow periodic deployment
of a tagged Am-Be neutron source at the equator of the steel shell. Gamma calibrations will be performed
with a tagged $^{22}$Na source moved through an additional circular tube installed vertically around the
steel shell.

The optical calibration systems consist of 20 LED-driven optical fibers for light injection, attached to
select lightguides. In addition to this, an internal diffuse and isotropic light source will
be used to characterize in detail the detector response at chosen wavelengths before and after the TPB
deposition, i.e. before cooling down the detector.

A more detailed technical paper describing the DEAP-3600 detector is in preparation.
\section{Background mitigation strategy}
The DEAP-3600 background budget predicts essentially a background free run, i.e. 0.6 background events
in the WIMP region of interest in 3 tonne-years from all sources~\cite{DEAPTaup2012}.

At approximately 1~Bq per kg of natural argon~\cite{Loosli83}, \textbeta/\textgamma\ from $^{39}$Ar will be
the dominant background, and pulse-shape discrimination (PSD) on the scintillation signal will be used to mitigate it, as described
in \cite{Boulay2006, Deap1Psd, PhysRevC.78.035801}. The effectiveness of PSD strongly depends on the light
yield (a factor of approximately 5$\times$10$^4$ difference between 4 and 8~pe/keV~\cite{DEAPTaup2012}).
High light yield requires very high photocathode coverage, which is best achieved in a single-phase detector with PMT coverage in
4\textpi\ (without argon vapor space and without high-voltage requirements/infrastructure needed to additionally extract information from the ionization channel).
Significant design efforts, guided by Monte Carlo simulations,
were devoted to reach the 8~pe/keV light yield needed to achieve the required level of PSD at
15~\kevee\ energy. The light collection was maximized through use of highly transparent acrylic and
through optimal choice and configuration of reflectors
(specular around the lightguides, as well as diffuse ones on the acrylic sphere between lightguides).

Cosmogenic backgrounds are mitigated by the depth of SNOLAB (overburden of 6000 m.w.e.) and the water \v{C}erenkov veto. 

In order to address the remaining neutron and radon-progeny-induced surface \textalpha\ backgrounds, we have engaged
in a material assay, selection and quality control campaign.
In addition to this we have performed substantial R\&D on low-radioactivity methods and PSD using the DEAP-1 detector,
which is described below.

\subsection{Selection and treatment of materials}
Extreme care has gone into the selection and handling of detector materials, to ensure that background target levels will be met.
All materials have been benchmarked against the design radiopurity targets using the germanium detector
at the SNOLAB Low Background Counting Facility~\cite{SnolabGE}
and using a low background $^{222}$Rn emanation setup at Queen's University.

Radon in the cryogen, which is of particular concern, is removed by
a custom-made charcoal radon trap installed in the argon supply lines (after 
initial purification of Ar gas with a commercial SAES getter). In order to reduce radon emanation
from surfaces exposed to the inner detector, we have carefully selected low-radioactivity welding electrodes
for use with the process systems lines, and used exclusively electropolished stainless steel tubing and
metal gaskets, ultrasonically cleaned and passivated with hot citric acid and then sealed hermetically. 

The fabrication of the AV and the lightguides was subject to extensive
quality control in co-operation with suppliers to ensure radiopurity of the bulk acrylic~\cite{JillingsLRT2013}.
Special care was taken to prevent radon from diffusing into the acrylic monomer where it would cause a buildup
of long-lived $^{210}$Pb, a source of backgrounds such as (\textalpha, n) neutrons~\cite{CaiLRT2010}.
Similar quality control was in place for other critical detector materials, including the polyethylene
of the neutron shielding blocks and the TPB used as wavelength shifter.

A dedicated system was developed to measure impurities in the bulk acrylic by vaporizing a large
quantity of acrylic and counting the concentrated residue with ultra-low background HPGe detectors
and a low background \textalpha\ spectrometer~\cite{NantaisLRT2013}. The measured upper limit for $^{210}$Pb
content is 10$^{-19}$ g/g~\cite{CorinaMSc}.

After completion of the AV it was sealed from the SNOLAB environment. While purged with low-radon gas
a few hundred micron layer was removed from the inner surface by a custom-made resurfacer robot.
Based on a model of Rn deposition and diffusion as well as exposure and thermal
history of the AV from its time of production through construction and annealing, we expect
to bring the surface radiopurity close to the level of the bulk acrylic and reduce the surface background
rate by more than an order of magnitude.

\subsection{DEAP-1 results}
DEAP-1 is a 7-kg liquid argon cell instrumented with two PMTs. After operating in the surface lab at Queen's,
it went through three major iterations
underground at SNOLAB. In the first generation, 5~inch ETL9390B (flat-face) PMTs
were used, and the light yield was about 2.7~pe/keV. A large dataset collected in that
configuration allowed us to demonstrate PSD on the level of $<$2.8$\times$10$^{-8}$ (Fig.~\ref{fig:deap1psd}) and
showed good agreement with the analytic PSD model, projecting PSD on the level of 10$^{-10}$ for DEAP-3600~\cite{Deap1Psd}.

For the other two iterations, the PMTs were replaced by Hamamatsu R5912-HQE PMTs -- the same type used for DEAP-3600 --
which resulted in a light yield increase to above 4~pe/keV, consistent with Monte Carlo predictions.
Data collected then was mostly used to study surface backgrounds, with a smaller amount of data taken to
demonstrate the feasibility of an enhanced PSD, based on Bayesian technique for photoelectron counting~\cite{Caldwell2014}.

In all cases the inner detectors were prepared in a radon-reduced glovebox, either using an inner-surface
sanding procedure akin to that employed for DEAP-3600 or with an alternative technique based on deposition
of purified acrylic coatings~\cite{KuzniakLRT2010}. 

Surface \textalpha\ background levels of approximately 100~\micro Bq/m$^2$ were reached in DEAP-1; this is
sufficiently low to achieve the target background level for the inner spherical AV surface in DEAP-3600.
The backgrounds were consistent with the presence of radon and its progeny in the detector
and longer lived radon daughters ($^{214}$Pb and/or $^{214}$Bi) plating out on the detector surface.
We successfully applied coincidence techniques, based on \textalpha-\textalpha\ tags ($^{222}$Rn-\/$^{218}$Po, $^{220}$Rn-\/$^{216}$Po)
and \textbeta-\textalpha\ tags ($^{214}$Bi-\/$^{214}$Po), to study and discriminate against those backgrounds~\cite{Amaudruz2014}.

As a by-product of this work, we developed a detailed simulation model of the inner detector and the interface between acrylic,
wavelength shifter and cryogen, including \textalpha-induced scintillation from TPB (measured also ex-situ, see~\cite{Pollmann2011}) and
realistic surface roughness effects~\cite{Amaudruz2014,Kuzniak2012} and found good agreement with the data.

The last generation of DEAP-1 was instrumented with a light injection system similar to the one installed in DEAP-3600, which was
used for accurate single photoelectron (SPE) charge calibration and resulted in an analytic model of the SPE distribution, based on
a superposition of two Gamma distributions.

Other technical solutions successfully prototyped in DEAP-1 and used in DEAP-3600 include: PMT mounts, silicon oil based PMT-lightguide
optical couplings, radon trap and MIDAS-based DAQ system.

\begin{center}
\begin{figure}[htb]
\begin{overpic}[width=\columnwidth,trim=0 0 1.5cm 0,clip=true]{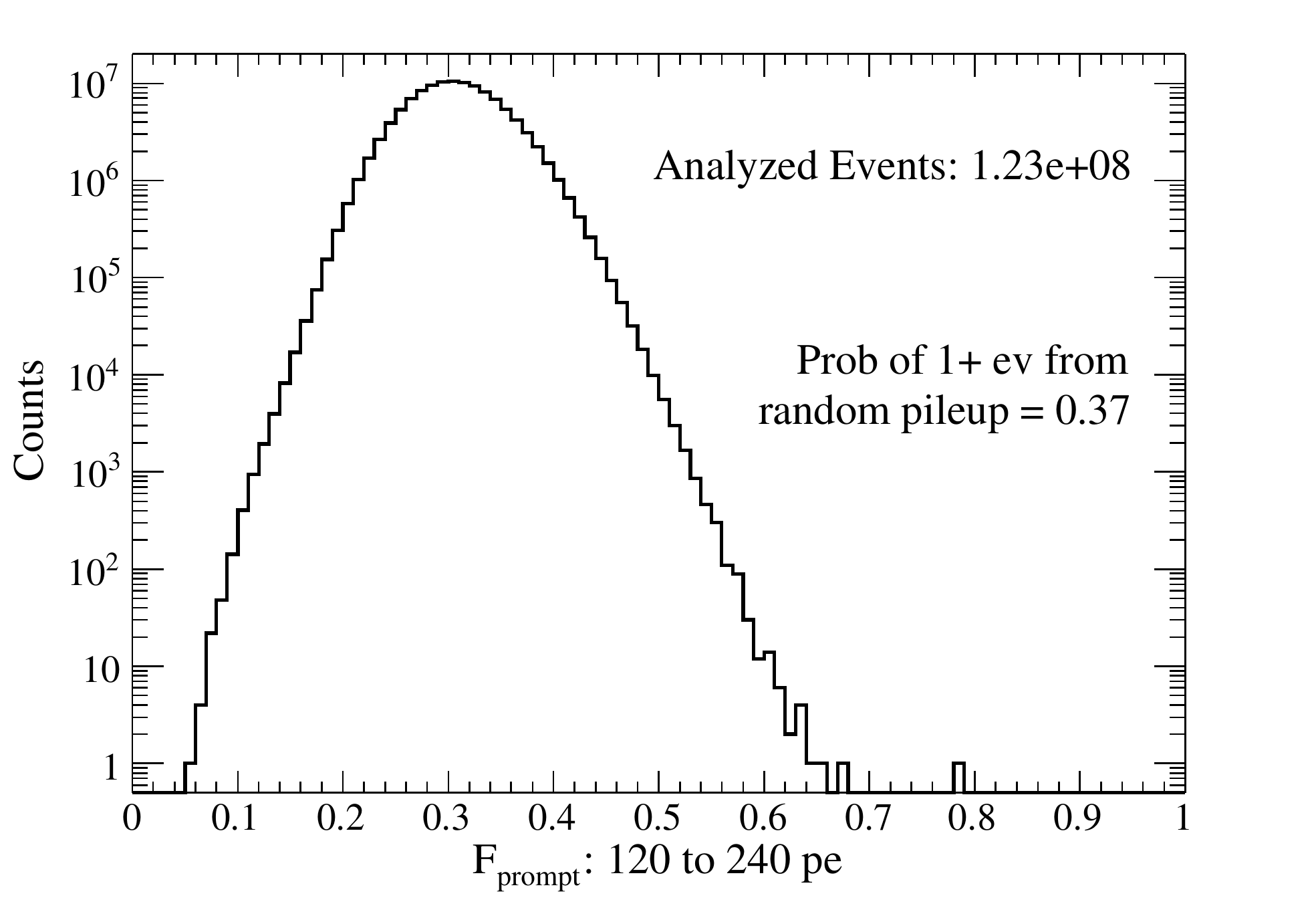}
\linethickness{0.5pt}
\multiput(71.4,11.5)(0, 2){31}{\color{red}\line(0,-1){1}}
\end{overpic}
\caption{The combined DEAP-1 PSD dataset containing tagged $^{22}$Na \textgamma\ events from surface and underground runs
at $\sim$2.7~pe/keV for 120-240~photoelectrons range~\cite{Deap1Psd}.
\fprompt\ is the pulse shape discrimination parameter used to separate electron and
nuclear recoils. There is only one event detected inside of the 90\%\ nuclear recoil acceptance window, i.e. above 0.7
(red dashed line), while the probability of detecting one or more background event from random pile-up is 0.37.
The resulting upper limit on electron recoil misidentification fraction
is $<$2.8$\times$10$^{-8}$ (90\%~C.L.) for 44-89~\kevee\ energy range.}
\label{fig:deap1psd}
\end{figure}
\begin{figure}[htb]
\includegraphics[width=\columnwidth]{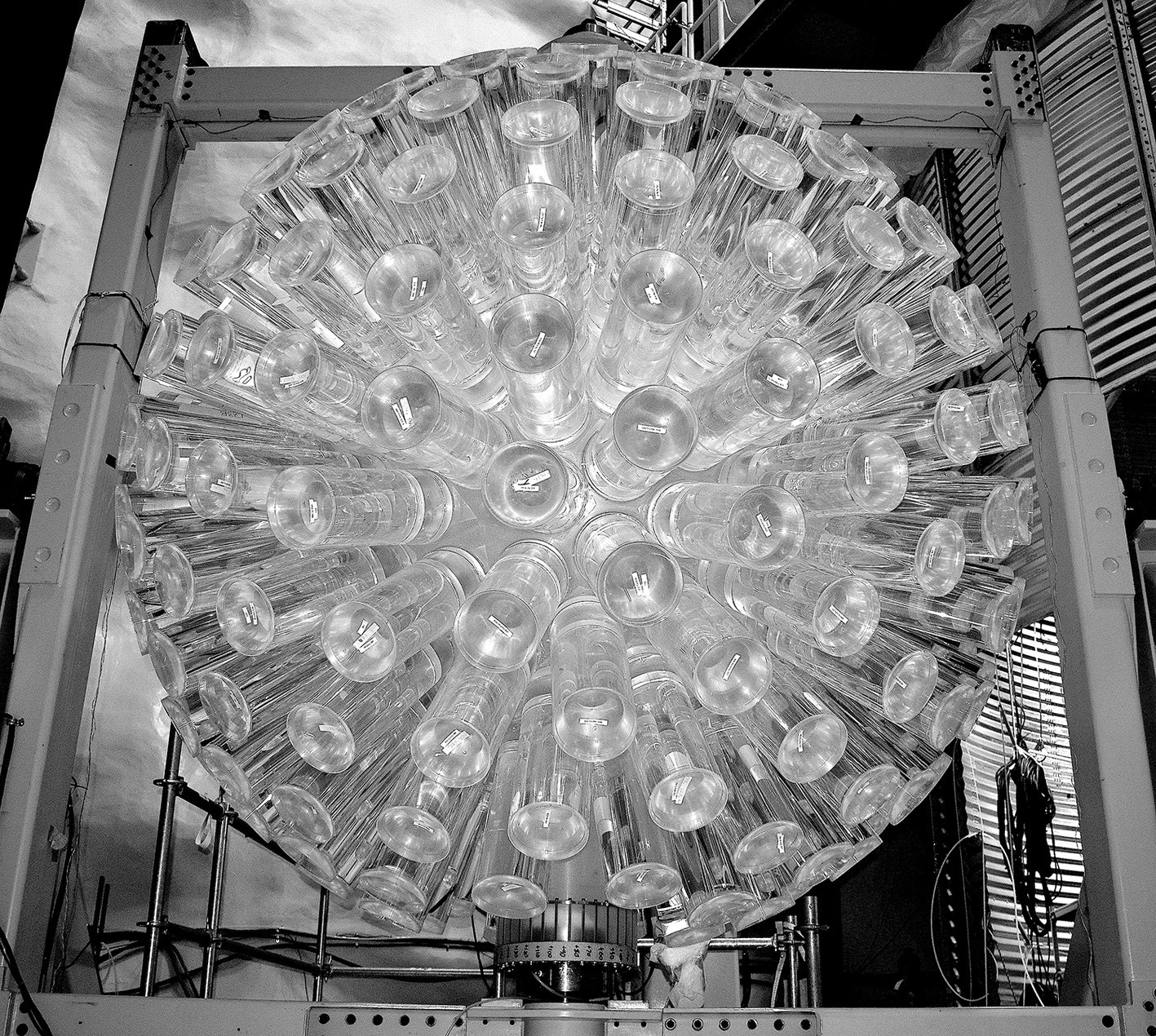}
\caption{The DEAP-3600 acrylic vessel attached to an assembly frame after completion of lightguide bonding. 
The full diameter with lightguides is about 3~m.}
\label{fig:av_photo}
\end{figure}
\begin{figure}
\includegraphics[width=\columnwidth]{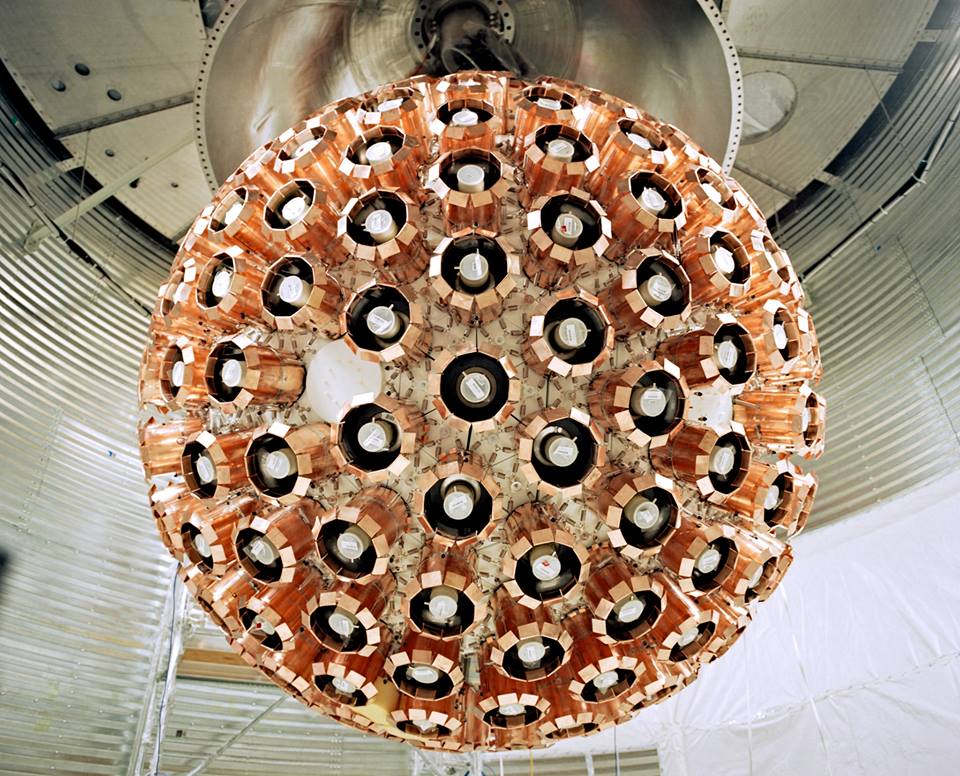}
\caption{The DEAP-3600 inner detector with installed PMTs, reflectors, neutron shielding polyethylene blocks and copper thermal shields.}
\label{fig:deap3600photo}
\end{figure}
\end{center}
\section{DEAP-3600 project status and timeline}
During 2013 the AV was fabricated and machined by Reynolds Polymer, Colorado, and the University of Alberta.
The final bonding of the AV and bonding of 255 acrylic lightguides onto the AV~\cite{McElroy2014} was performed underground and included
multiple high-temperature anneals in low-radon atmosphere (see Fig.~\ref{fig:av_photo}).

In early 2014 the inner detector was dressed in diffuse and specular reflectors and polyethylene shielding blocks as well as
instrumented with magnetically and thermally shielded PMTs, temperature sensors and
light injection ports (Fig.~\ref{fig:deap3600photo}). Other
activities, such as installation and commissioning of the process systems, electronics and calibration systems, were happening in parallel. 

After resurfacing of the AV is finished (ongoing), the AV will be sealed inside
the steel shell and the TPB coating will be applied. First 'dark' PMT and optical calibration data will also be taken at that time.

The detector is currently entering the commissioning phase. A short run with gaseous argon is planned before and
during the detector cooldown for initial assessment of detector response and backgrounds.
DEAP-3600 will reach its final projected sensitivity to the spin-independent WIMP-nucleon cross-section of 10$^{-46}$~cm$^2$ after three years of running.
For WIMP masses larger than about 100~GeV it will become competitive with current best limits after collecting two months' worth of data.
\begin{center}
\begin{figure}[htb]
\includegraphics[width=\columnwidth,trim=6cm 0 3cm 0,clip=true]{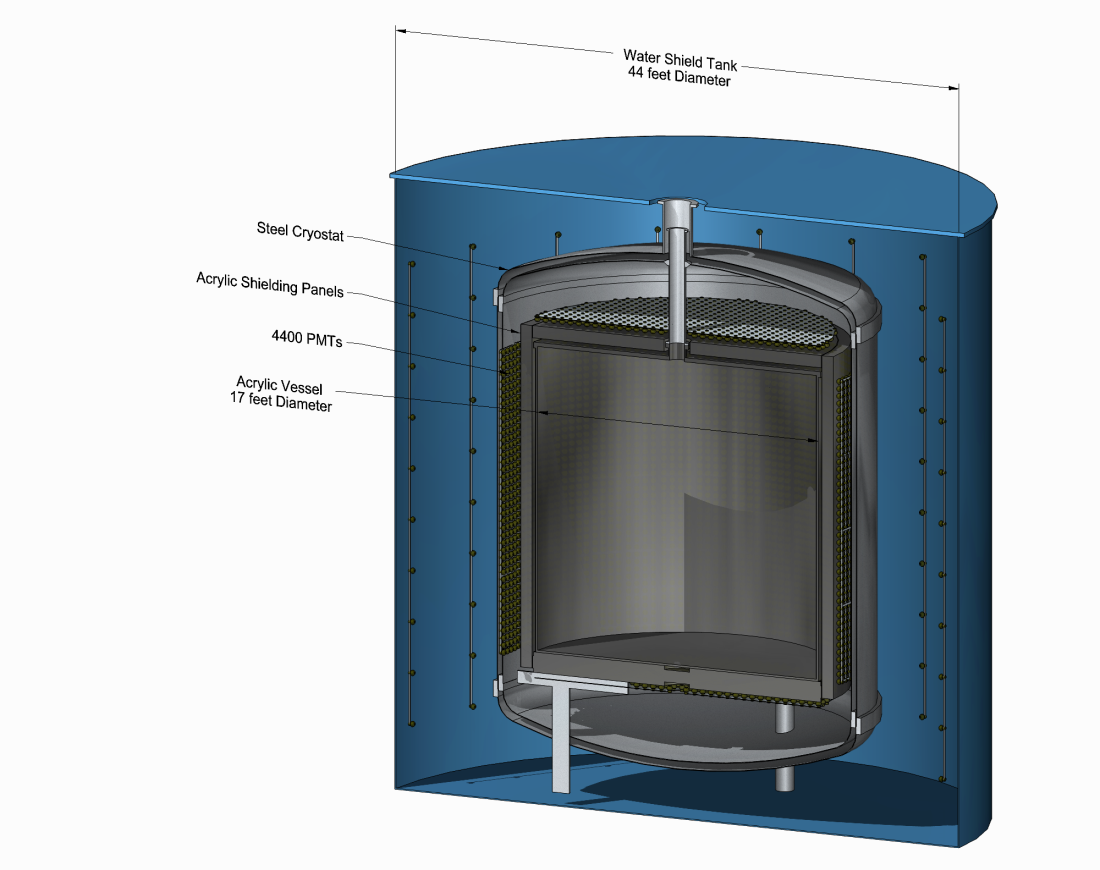}
\caption{Concept for 50-tonne liquid argon detector. The inner acrylic vessel is a conventional
cylindrical design with flanged lid, approximately 17~feet diameter, and is sanded for control of
radon surface contamination. Acrylic shielding panels provide neutron moderation from the 4400
high-QE PMTs. The outer shielding tank is approximately 43-feet in diameter.}
\label{fig:deap-50t}
\end{figure}
\end{center}

\section{Future scale up to DEAP-50T}
Detectors based on the single-phase technology can be scaled up to very large masses with relative ease,
as extraction of information requires only reading out scintillation photons.

The conceptual design for a large argon detector with sensitivity to a spin-independent WIMP scattering
cross-section per nucleon of 2$\times$10$^{-48}$~cm$^2$ is based on a fiducial mass of 50-tonnes
with a low-background 3-year exposure (Fig.~\ref{fig:deap-50t}). Although the final detector
design will be optimized based on detailed Monte-Carlo simulations,
the basic concept has been informed by development of single-phase argon detectors, and
requires no significant additional R\&D, other than scaling the size of the vessel and the supporting
detector systems. At this larger scale, impurity constraints from surface backgrounds,
which are a dominant concern for the smaller detectors, are relaxed since the significantly larger detector
more easily allows maintaining a significantly higher fiducial volume while mitigating external
backgrounds with vertex reconstruction.
The diameter of the detector presented here is approximately
three times the diameter of the DEAP-3600 detector, and so the detector becomes large
enough to make better use of timing information for vertex reconstruction, without significantly affecting
the ability to detect photons.

The detector consists of a large inner acrylic vessel, which serves to seal the inner argon volume
from any sources of radon once the detector has been constructed. For
ease of construction, this vessel is designed as a large cylinder with a removable lid.
It is surrounded by approximately 12-inches of clear and ultralow-background
acrylic panels to provide required neutron shielding from the PMTs. This acrylic and the PMTs
themselves are immersed in a buffer volume of liquid argon, so that there are no significant thermal
gradients on the acrylic vessel - its purpose is only for control of radon emanation into the active
detector region. The inner assembly is contained in a double-walled vacuum cryostat, which itself is immersed
in the large water shielding tank. The nominal design parameters are shown in Table~\ref{tab:deap-50t}.
\begin{table}
\centering
\small
   \begin{tabular}{@{} l|l @{}} 
 \toprule
      Parameter & Value  \\
 \midrule
AV diameter & 5.2 m (17 feet) \\
Position resolution & 15 cm \\
Surface backgrounds & 100 per m$^2$ per day \\
No. PMTs & 4400 \\
Acrylic shield panels & 12-inch thick \\
Steel Shell diameter & 7.1 m (24 feet) \\
Shield tank diameter & 13.4 m (44 feet) \\
Argon in AV & 150 tonnes \\
Argon buffer & 240 tonnes  \\
 \bottomrule
\end{tabular}
\caption{Nominal design parameters for 50-tonne argon detector with spin-independent WIMP-nucleon sensitivity of 2$\times$10$^{-48}$~cm$^2$.
Low-radioactivity argon is held in a cylindrical acrylic tank with a lid, instrumented with 8-inch PMTs. Surface background rate corresponds to 3-month exposure to ambient air.}
\label{tab:deap-50t}
\end{table}

For the large detector proposed here, some level of depletion of $^{39}$Ar will be required,
in order to limit the pileup rate, and in order for PSD to remove the background \textbeta\ events to a sufficiently low level.
Recent work has shown that there is a possibility of obtaining argon that has been sequestered underground
and is depleted in $^{39}$Ar by a factor of 100~\cite{Back2012}.
This depletion factor is good enough to bring pileup and overall rates to an acceptable level
for target masses up to several hundred tonnes of argon. Obtaining approximately 150 tonnes of low-radioactivity
argon for the experiment will require significant effort.

\subsection{Physics reach}
\begin{center}
\begin{figure}
\includegraphics[width=\columnwidth]{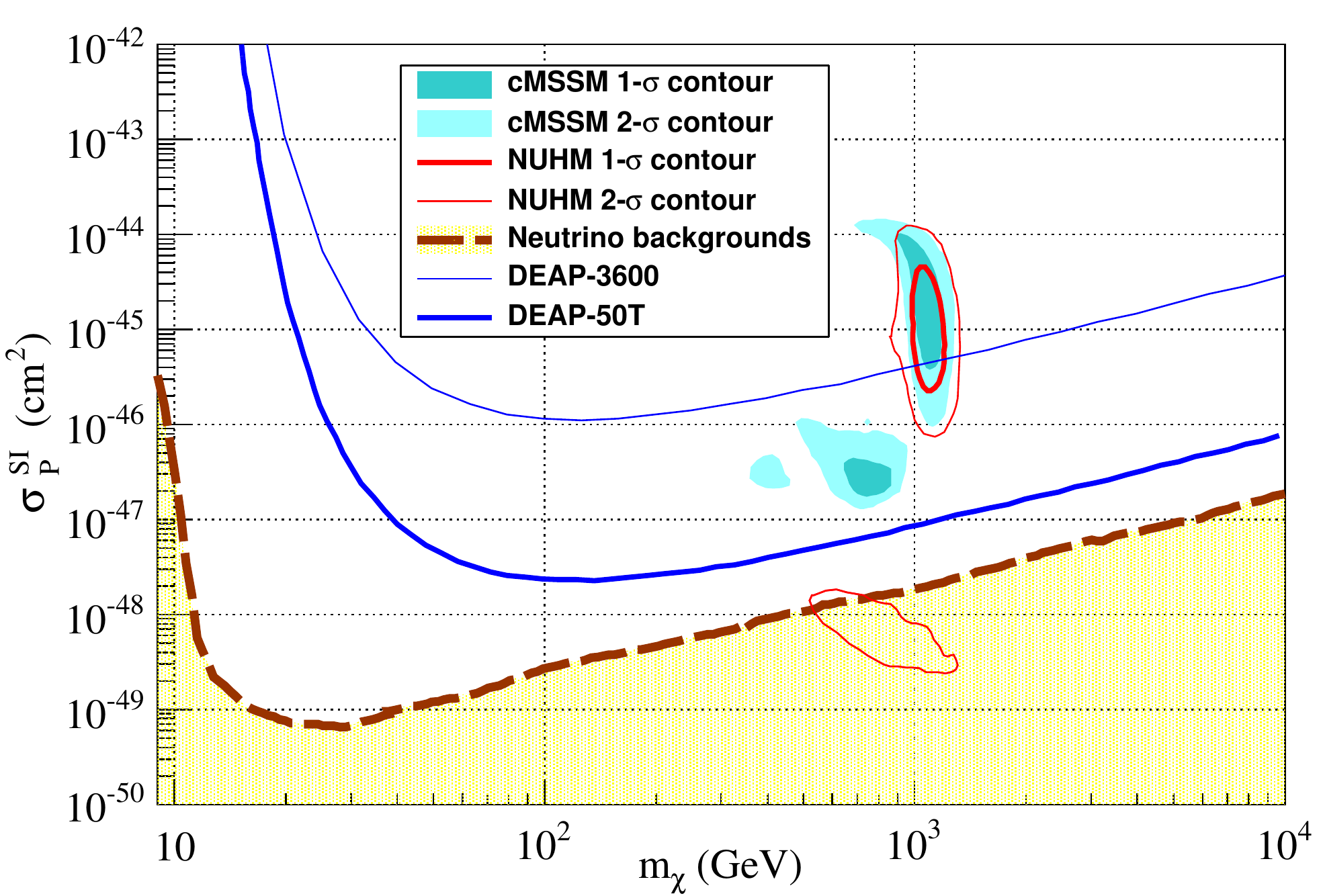}
\caption{Exclusion sensitivity (90\%~C.L.) of the proposed DEAP-50T argon detector compared with the sensitivity of DEAP-3600. 
The 1-\textsigma\ and 2-\textsigma\ contours are shown for NUHM and cMSSM models, as adapted from Ref.~\cite{Roszkowski2014}.
The approximate ultimate limit for the WIMP discovery potential of non-directional searches is represented by
the thick dashed line~\cite{Billard2014} (in the shown energy range mostly driven by atmospheric and diffuse supernova neutrinos).}
\label{fig:deap-50t-phys}
\end{figure}
\end{center}

A large argon experiment with sensitivity to WIMPs at the 2$\times$10$^{-48}$~cm$^2$ level will be capable of probing the full parameter space
of the most-studied supersymmetric models that are consistent with the current global data~\cite{Roszkowski2014}. The
experimental sensitivity (using only information from the 50-tonne detector) is shown in Fig.~\ref{fig:deap-50t-phys}; in
addition to the possibility of a positive detection, the detector will allow setting strong constraints on WIMP properties, including mass.

At the above sensitivity levels, neutrinos from various sources can become the limiting
background for liquid noble gas detectors. 
At low recoil energies, the dominant source of neutrino background events in both
xenon and argon is from coherent neutrino-nuclear scattering, which leads to a recoiling nucleus
in the experiment that cannot be discriminated against.
Additionally, elastic scattering of electron neutrinos from the Sun on electrons~\cite{Strigari2009} is a 
particular concern for xenon. Assuming the nominal electron recoil discrimination of 0.995 achieved 
by XENON100~\cite{Aprile2012}, the pp and $^{7}$Be solar neutrinos are the dominant neutrino background
above threshold, generating approximately 0.15 events per tonne-year of xenon exposure,
in the assumed analysis window of 2-10~\kevee~\cite{Baudis2014}. 
Argon has essentially no sensitivity to solar neutrino backgrounds,
since pulse-shape discrimination of electron recoils is many orders of magnitude more powerful than
in xenon. The solar neutrino backgrounds limit the practically achievable
sensitivity to scattering of heavy dark matter particles ($\sim$100~GeV) to approximately $\sim$10$^{-47}$~cm$^2$
for xenon, while argon would be ultimately limited by coherent scattering of atmospheric neutrinos, at
a cross-section sensitivity below 10$^{-48}$~cm$^2$.

\section{Summary}
DEAP-3600 is a single-phase liquid argon detector with projected sensitivity to the WIMP-nucleon cross-section of 10$^{-46}$~cm$^2$.
At the light yield of 8~pe/keV and with natural argon DEAP-3600 will run at 15~\kevee\ threshold, limited by the achievable
pulse shape-discrimination for $^{39}$Ar \textbeta's. Stringent radiopurity targets have been defined based on Monte Carlo simulations
in order to meet the goal of a background-free 3 year run. We have invested significant efforts and R\&D into background mitigation,
sourcing radiopure materials, and implementing low-radioactivity techniques, including limiting
the exposure to radon at all steps of the detector production and assembly. DEAP-1,
a 7-kg liquid Ar prototype detector, was used to test a number of technical solutions that were applied in DEAP-3600,
and to demonstrate the potential for low backgrounds. The final steps of the DEAP-3600 construction are ongoing, in parallel with commissioning 
of the critical detector subsystems. DEAP-3600 will start running at the beginning 2015, and will reach a sensitivity competitive with
the best current limits after a few months of data taking.

For the future, we are developing the concept of a large scale detector, DEAP-50T, 
with 50 tonnes fiducial mass of low radioactivity argon. Such a detector will conclusively probe most of the remaining parameter
space of the leading and simplest supersymmetry models (cMSSM and NUHM), which largely remain beyond the reach of the LHC, and
can severely constrain more general models. 

\section*{Acknowledgements}
This work is supported by the National Science and Engineering Research Council of Canada (NSERC), by the Canada Foundation for Innovation (CFI), by the Ontario Ministry of Research and Innovation (MRI) and by the European Research Council (ERC). We thank Compute Canada, Calcul Qu\'ebec and the High Performance Computing Virtual Laboratory (HPCVL), Queen's University site, for computational support and data storage.  We are grateful to SNOLAB staff for excellent on-site support.



\bibliographystyle{elsarticle-num}
\bibliography{ichep2014_mk_arxiv}

\begin{thebibliography}{10}
\expandafter\ifx\csname url\endcsname\relax
  \def\url#1{\texttt{#1}}\fi
\expandafter\ifx\csname urlprefix\endcsname\relax\def\urlprefix{URL }\fi
\expandafter\ifx\csname href\endcsname\relax
  \def\href#1#2{#2} \def\path#1{#1}\fi

\bibitem{WMAP}
D.~N. Spergel~et al., ApJS 148 (2003) 175.

\bibitem{Planck}
P.~Ade et~al. (Planck), {A\&A (2014), }\href
  {http://dx.doi.org/10.1051/0004-6361/201321591}
  {\path{doi:10.1051/0004-6361/201321591}}.

\bibitem{DEAPTaup2012}
M.~G. Boulay~(DEAP), J. Phys.: Conf. Ser. 375 (2012) 012027.

\bibitem{Gorel2014}
P.~Gorel~(DEAP), {2014, }\href {http://arxiv.org/abs/1406.0462}
  {\path{arXiv:1406.0462}}.

\bibitem{Akerib2014}
D.~Akerib et~al. (LUX), Phys. Rev. Lett. 112 (2014) 091303.

\bibitem{Roszkowski2014}
L.~Roszkowski~et al., JHEP 08 (2014) 67.

\bibitem{Kowalska2013}
K.~Kowalska~et al., JHEP 06 (2013) 78.

\bibitem{Strege2013}
C.~Strege~et al., JCAP 04 (2013) 013.

\bibitem{Bechtle2012}
P.~Bechtle~et al., JHEP 06 (2012) 098.

\bibitem{Xenon1T}
E.~Aprile et~al. (XENON1T), {2012, }\href {http://arxiv.org/abs/1206.6288}
  {\path{arXiv:1206.6288}}.

\bibitem{Fowlie2013}
A.~Fowlie~et al., Phys. Rev. D 88 (2013) 055012.

\bibitem{cdms}
{CDMS collaboration}, Science 327 (2010) 1619.

\bibitem{Aprile2012}
E.~Aprile et~al. (XENON), Phys. Rev. Lett. 109 (2012) 181301.

\bibitem{Pollmann2012}
T.~Pollmann~(DEAP), Physics in Canada 68 (2012) 142.

\bibitem{Amaudruz2012}
P.-A. Amaudruz~et al., in: Real Time Conference (RT), 2012 18th IEEE-NPSS,
  2012, pp. 1--8.
\newblock \href {http://dx.doi.org/10.1109/RTC.2012.6418165}
  {\path{doi:10.1109/RTC.2012.6418165}}.

\bibitem{Retiere2012}
F.~Retiere~et al., in: Nuclear Science Symposium and Medical Imaging Conference
  (NSS/MIC), 2012 IEEE, 2012, pp. 1802--1805.
\newblock \href {http://dx.doi.org/10.1109/NSSMIC.2012.6551421}
  {\path{doi:10.1109/NSSMIC.2012.6551421}}.

\bibitem{Loosli83}
H.~H. Loosli, Earth and Planetary Sci. Lett. 63 (1983) 51.

\bibitem{Boulay2006}
M.~Boulay, A.~Hime, Astropart. Phys. 25 (2006) 179.

\bibitem{Deap1Psd}
{P.-A. Amaudruz et al. (DEAP)}, {{to be published in Astropart. Phys. (2014),
  }}\href {http://arxiv.org/abs/0904.2930} {\path{arXiv:0904.2930}}.

\bibitem{PhysRevC.78.035801}
W.~H. Lippincott~et al., Phys. Rev. C 78 (2008) 035801.

\bibitem{SnolabGE}
I.~Lawson, B.~Cleveland, {AIP} Conf. Proc. 1338 (2011) 68.

\bibitem{JillingsLRT2013}
C.~Jillings~(DEAP), AIP Conf. Proc. 1549 (2013) 86.

\bibitem{CaiLRT2010}
B.~Cai, M.~Boulay, B.~Cleveland, T.~Pollmann~(DEAP), AIP Conf. Proc. 1338
  (2011) 137.

\bibitem{NantaisLRT2013}
C.~M. Nantais, B.~T. Cleveland, M.~G. Boulay, AIP Conf. Proc. 1549 (2013) 185.

\bibitem{CorinaMSc}
C.~M. Nantais, Master's thesis, Queen's University (2014).

\bibitem{Caldwell2014}
M.~Akashi-Ronquest~et al., {2014, }\href {http://arxiv.org/abs/1408.1914}
  {\path{arXiv:1408.1914}}.

\bibitem{KuzniakLRT2010}
M.~Ku\'zniak~(DEAP), AIP Conf. Proc. 1338 (2011) 101.

\bibitem{Amaudruz2014}
{P.-A. Amaudruz et al. (DEAP)}, {Astropart. Phys.} 62 (2015) 178.

\bibitem{Pollmann2011}
T.~Pollmann, M.~Boulay, M.~Ku\'zniak, NIM A 635 (2011) 127.

\bibitem{Kuzniak2012}
M.~Ku{\'z}niak, M.~G. Boulay, T.~Pollmann, Astropart. Phys. 36 (2012) 77.

\bibitem{McElroy2014}
T.~McElroy~(DEAP), Physics in Canada 70 (2014) xx.

\bibitem{Back2012}
H.~O. Back~et al., {2012, }\href {http://arxiv.org/abs/1204.6024}
  {\path{arXiv:1204.6024}}.

\bibitem{Strigari2009}
L.~E. Strigari, New J. Phys. 11 (2009) 105011.

\bibitem{Baudis2014}
L.~Baudis~et al., JCAP 01 (2014) 044.

\bibitem{Billard2014}
J.~Billard~et al., Phys. Rev. D 89 (2014) 023524.

\end{thebibliography}







\end{document}